\documentclass[usegraphicx,useAMS,usenatbib]{mn2e}
\usepackage{aas_macros}
\usepackage{amssymb}
\title[The density profile of DM haloes beyond the virial radius]{The dark outside: the density profile of dark matter haloes beyond the virial radius}
\author[H. Tav{\'\i}o et al.]{Hugo Tav{\'\i}o,$^{1,2,3}$\thanks{E-mail: htavio@iaa.es} Antonio J. Cuesta$^1$, Francisco Prada,$^1$ Anatoly A. Klypin$^{4}$,\newauthor Miguel A. S\'anchez-Conde$^1$\\
$^1$Instituto de Astrof{\'\i}sica de Andaluc{\'\i}a (CSIC),
Camino Bajo de Hu\'etor 50, E-18008 Granada, Spain. \\
$^2$Instituto de Ense\~{n}anza Secundaria "Jos\'e Cadalso", c/Castiella s/n, San Roque (C\'adiz), Spain.\\
$^3$Escuela Polit\'enica Superior de Algeciras. Universidad de C\'adiz, Avda. Ram\'on Pujol s/n 11202 Algeciras (C\'adiz), Spain.\\
$^4$Astronomy Department, New Mexico State University, MSC 4500, P.O.Box 30001, Las Cruces, NM,
880003-8001, USA.}

\begin{document}

\bibliographystyle{mn2e}

\begin{small}\end{small}\maketitle

\begin{abstract}

We present an approximation for the average density profile of dark matter haloes in the $\Lambda$CDM cosmological model, which is accurate to within 10--15\% even for large radial distances from $0.05R_{\rmn {vir}}$ up to $10R_{\rmn {vir}}$ for halo masses ranging from $10^{11.5}$ to $10^{15.0}$ $h^{-1}M_{\odot}$. We propose a modified form of the Navarro, Frenk \& White (NFW) approximation: $\rho(r)=\rho_{\rmn{NFW}}(r)+ A (r/R_{\rmn {vir}})^{-1}+B (1+r/R_{\rmn {vir}})^{-1}$. This generalized expression, which is applicable to the external regions of dark matter haloes, only very slightly affects the density in the inner regions of haloes. The strong correlation among the different parameters in the model allows us to describe the profile in terms of just one parameter: the virial mass. We integrate our density profile to derive the enclosed mass in a sphere of a given radius and compare it with the NFW results. We find that the NFW underestimates the enclosed mass by more than 50\% at $10R_{\rmn {vir}}$, whereas our model reproduces the results from numerical simulations to within 2\% accuracy even at this distance. We also use this new approximation to study the weak gravitational lensing and to obtain an analytic expression for the tangential shear. This allows us to quantify the contribution to the shear from the outer regions of the density profile. For the first time we calculate the difference between the tangential shear calculated via the NFW profile and the corresponding result when the external regions of haloes in cosmological simulations are taken into account. We find a 4\% difference for all the mass ranges under study.

\end{abstract}


\section{Introduction}

\label{sec:intro}

Measurements of the mass distribution of dark matter (DM) haloes are essential for the theory of structure formation. This issue has been addressed in two ways: observational and theoretical. The observational measurements have profited from recent gravitational lensing studies and also from measurements of galactic dynamics using large scale and massive galaxy surveys (e.g. \citealt{Ma06}, \citealt{Pr03}). Theoretical estimates made substantial progress thanks to recent advances in both top-hat gravitational collapse models and fully non-linear evolution via $N$-body simulations (e.g. \citealt{Sa07}, \citealt{As07}, \citealt{Di07}).

Cosmological $N$-body simulations have been extensively used in order to provide predictions for the structure of DM haloes in the hierarchical clustering scenario. The Navarro, Frenk \& White (NFW) density profile has become a convenient formula for the description of DM haloes in a broad mass range. This profile provides a reasonable fit for the halo density distribution, despite that many other functions have been suggested in order to improve accuracy (\citealt{Mo98}, \citealt{Se68}, \citealt{Ei65} profiles, see also \citealt{Me05} for reference). Nevertheless, the NFW density profile \citep{Na97} was proposed and contrasted to fit the inner regions of DM haloes roughly up to the virial radius, where haloes are expected to be in equilibrium \citep{Co96}. Therefore, it was not expected that the NFW analytical profile will describe the external density profile properly. Our aim here is to extend the validity of this approximation to describe the outer regions of DM haloes. Moreover, recent works (see \citealt{Pr06}, \citealt{Cu07}, \citealt{Di07}) have shown that the DM virialized regions extend well beyond the formal virial radius especially for low-mass haloes. This reinforces our interest in studying haloes beyond the virial radius.

On the other hand, current observational techniques allow us to measure the mass distribution around galaxies and clusters at large distances. The dynamics of satellite galaxies (e.g. \citealt{Za94}; \citealt{Za97}; \citealt{Pr03}; \citealt{Br04}; \citealt{Co07}) provide strong constraints on the shape of the density profiles of DM haloes in the outer regions. Another technique which has proven its usefulness despite its recent development, is the analysis of the weak lensing effect in gravitational lensing theory (e.g. \citealt{Me99}; \citealt{Ba01}; \citealt{Ho04} \citealt{He06}; \citealt{Mas07}). In both observational techniques, the external regions under study are often beyond the virial radius. Yet, the use of the NFW profile to model observations is still very common in the determination of DM halo mass. Although the NFW profile is a great improvement as compared to the isothermal profile to describe for example the weak lensing observations (e.g. \citealt{Ma06b}), it is not suitable for extrapolation at large distances beyond the virial radius. The study of the physical processes occurring in the external regions of DM haloes will provide unique insight into our understanding of the formation of DM haloes. A good example is the recent observational work by \citealt{Ge08} who measured temperature and density profiles beyond the virial radius of a cluster and this allowed them to produce improved constraints on the mass and gas fraction profiles. A first step in this direction is the study of average density profiles in the outer regions of DM haloes in cosmological simulations.

In this paper, we present an approximation for the density profile of DM haloes well beyond the virial radius, obtained from high resolution cosmological simulations. This approximation focuses on the external regions of the haloes as an extension of the NFW formula that fits well the inner regions. We took the NFW model as our starting point because of its simplicity and with the aim to derive an expression which is easy to implement in many of the applications mentioned above. In order to describe representative DM halo density profiles and compare them with our approximation, we average over many individual density profiles for a given mass range. Hence, our conclusions apply for the ensemble of haloes of a given mass, with a halo-to-halo mass dependent scatter. As it is already known from the early work of \citet{Na96}, individual haloes at a given range of mass usually show a similar profile. Thus, the analysis of average halo density profiles happens to be a reasonable option for our purposes.

The approximation presented in this paper includes free parameters which have to be fixed for a given mass range. However, we will also study some correlations among them and prove that only one parameter (for example the virial mass) is enough in order to provide a high quality fitting function for the inner and outer regions. Moreover, our approximation can be useful to obtain a suitable formula to describe the projected mass density of average DM haloes. We determine with this formula the tangential shear associated to gravitational lensing. We also have compared our results with those obtained using the NFW formula, in order to assess its validity at large distances from the halo centre.

The paper is organized as follows. In Section 2, we introduce our approximation that describes the density profiles of DM haloes well beyond $R_{\rmn{vir}}$. In Section 3, we fit the numerical density profiles drawn from cosmological simulations. In Section 4, we show the application of our model to the measurement of the tangential shear in gravitational lensing. Finally, we discuss and summarize the main conclusions in Section 5.

\section{The theoretical framework}
\label{sec:methods}

Let us start with a brief review of the NFW approximation for DM haloes in cosmological $N$-body simulations (see also e.g. \citealt{Lo01}). This is a simple function which provides a good description for the density profile inside the virial radius. Yet, it turns out to give wrong results for the external regions since it does not take into account the mean matter density of the Universe, which is the main contribution at large distances from the halo centre (see \citealt{Pr06}). The analytic expression for the NFW density profile is:

\begin{equation}
\rho_{\rmn{NFW}}(r)=\frac{\rho_{s}^{\rmn{NFW}}}{(r/r_{s})\left(1+r/r_{s}\right)^2}.
\label{eq:1}
\end{equation}

\begin{figure}
\includegraphics[width=0.5\textwidth]{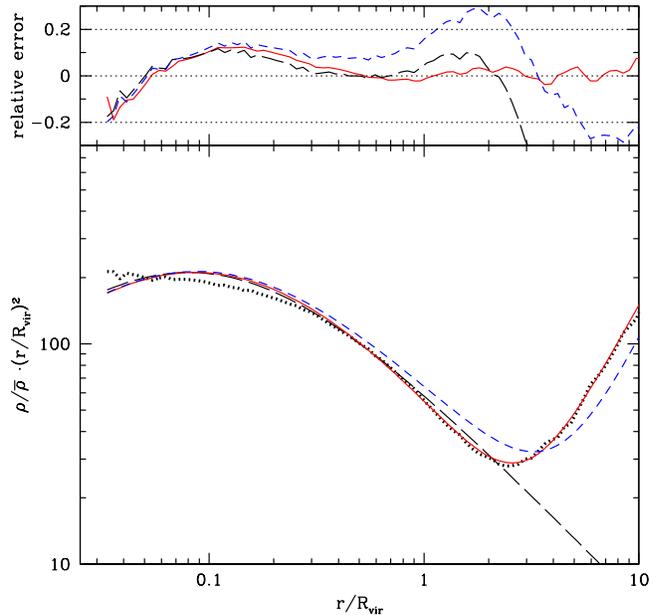}
\caption{Average matter density profile for a Milky Way size DM halo. The numerical profile given by cosmological simulations (dotted line) is compared to the fit to the NFW model (Eq.~\ref{eq:nfw}, long dashed line), the mean matter density of the Universe (Eq.~\ref{eq:nfwmod}, short dashed line), and our model (Eq.~\ref{eq:ourmodel}, solid line). The upper panel shows the relative error for each approximation with respect to the numerical profile.}
%\label{fig:1}
\end{figure}

This formula describes the internal profile region quite well (up to one virial radius approximately) and depends on two parameters: the characteristic density $\rho_{s}^{\rmn{NFW}} $ and the characteristic scale radius $r_{s} $. The latter is the radius in which $r^2\cdot\rho_{\rmn{NFW}}(r) $ reaches its local maximum, i.e. $ \frac{d \ \ln\rho(r)}{d \ln r}=-2$, and the former is just given by $\rho_{s}^{\rmn{NFW}}=4\rho_{\rmn{NFW}}(r_s)$. The NFW profile may also be described in terms of the virial mass $M_{\rmn{vir}}=M(<R_{\rmn{vir}}) $, and the concentration parameter $c \equiv R_{\rmn{vir}}/r_{s} $. The quantity $R_{\rmn{vir}}$ is the virial radius of the halo, defined as the radius of a sphere enclosing a given overdensity, the value adopted here is $\Delta=340$. Hence, the relation between the virial radius and the virial mass is $M_{\rmn{vir}}=\frac{4\pi}{3} 340 \ \bar{\rho} \ R_{\rmn{vir}}^3$. The NFW profile can now be written as follows:

\begin{equation}
\rho_{\rmn{NFW}}(s)=\frac{\rho_{s}^{\rmn{NFW}}}{cs\left(1+cs\right)^2}.
\label{eq:nfw}
\end{equation}

where $s$ is the radial coordinate scaled to $R_{\rmn{vir}}$, i.e. $s \equiv r/R_{\rmn{vir}}$. Note that $\rho_{s}^{\rmn{NFW}}$ is now a function (in general) of both $c$ and $M_{\rmn{vir}}$. In order to find this function, let us calculate the halo enclosed mass from the density NFW profile:

\begin{equation}
M_{\rmn{NFW}}(s)=\frac{3M_{\rmn{vir}}}{340\bar{\rho}}\frac{\rho_{s}^{\rmn{NFW}}}{c^3}\left[ \ {\ln(cs+1)-\frac{cs}{cs+1}} \right].
\end{equation}

Since $M_{\rmn{vir}}$ is defined as the mass inside the virial radius, we have,

\begin{equation}
M(s=1)\equiv M_{\rmn{vir}}
\label{eq:mv1}
\end{equation}

It is straightforward to derive the expression for the characteristic density in terms of $c$ and $M_{\rmn{vir}}$. It turns out that $\rho_{s}^{\rmn{NFW}}$ is actually a function of the concentration parameter only, i.e.:

\begin{equation}
\rho_{s}^{\rmn{NFW}}=\frac{340}{3}\,c^3g(c)\bar{\rho}\\
\label{eq:rhosnfw}
\end{equation}
where the function $g$ is defined as follows:
\begin{equation}
g(x)\equiv \frac {1} {\ln(x+1)-x/(x+1)} \nonumber\\
\end{equation}
Therefore, the mass inside a sphere of a given radius is completely given by the virial mass of the halo and its concentration:
\begin{equation}
\frac {M(s)}{M_{\rmn{vir}}}=g(c) \left[\ln(cs+1)-\frac {cs}{cs+1} \right]=\frac{g(c)}{g(cs)}.
\end{equation}

To describe the external regions of DM haloes, it is essential to decide the kind of function that must be added to the NFW approximation, in such a way that its contribution can be neglected in the inner parts. A first attempt to improve this fit is proposed in \citet{Pr06}. At very large radii the DM density profile should not tend to zero as the NFW profile does, but instead it should tend to the mean matter density of the Universe. Then, the modified profile is:

\begin{equation}
\rho_{\rmn{mod}}(s)=\frac{\rho_{s}^{\rmn{mod}}}{cs\left(1+cs\right)^2}+\bar{\rho}
\label{eq:nfwmod}
\end{equation}
The parameter $\rho_{s}^{\rmn{mod}}$ is no longer given by Eq.~(5). If we use the condition (4) we obtain in this case:

\begin{equation}
\rho_{s}^{\rmn{mod}}=113\,c^3g(c)\bar{\rho}=\rho_{s}^{\rmn{NFW}}-\frac {1} {3}c^3g(c)\bar{\rho}
\end{equation}

In addition, cosmological DM halo density profiles show a tiny or null influence on the concentration parameter at large distances ($r>R_{\rmn{vir}}$), so that the density profile in the halo outskirts can only depend on the remaining parameter $M_{\rmn{vir}}$. Therefore, in order to approximate the transition to the outer regions we add a function $f$ which depends only on $s$ (i.e. on the virial radius, but not on the concentration). This makes $M_{\rmn{vir}}$ the only relevant parameter at large distances from the halo centre. The new density profile approximation that we propose here is a simple extension of the NFW formula, i.e.,
\begin{equation}
\rho(s)=\frac{\rho_{s}}{cs\left(1+cs\right)^2}+\bar{\rho}+f(s)\bar{\rho}
\end{equation}
where $f(s)$ is the new function that we are proposing in order to improve the fit to the external parts of DM haloes. We choose the following rational function:
\begin{equation}
f(s)=\frac{b_{1}} {s}+\frac{b_{2}} {s+1}
\label{eq:13b}
\end{equation}
There are several reasons for this choice. On the one hand, we aim for a simple formula which allows an easy calculation for the tangential shear. On the other hand, we want to minimize the influence of the new added parameters ($b_1$ and $b_2$) on the internal profile, where the NFW profile is steeper, except for the very inner regions in which NFW and $f(s)$ are both proportional to $r^{-1}$. We thus propose in this work the following model as an approximation of the density profile from the inner regions up to large distances well beyond $R_{\rmn{vir}}$:

\begin{equation}
\rho(s)=\frac{\rho_{s}}{cs\left(1+cs\right)^2}+\left( \frac{b_{1}}{s}+\frac{b_{2}}{s+1}+1 \right)\bar{\rho}
\label{eq:ourmodel}
\end{equation}
It is worth noting that the enclosed mass and the characteristic density are now given by these relations:

\begin{eqnarray}
   M(s)&=& \frac{3 \, M_{\rmn{vir}}}{340} \left[ \frac{\rho_s}{c^3g(cs) \bar{\rho}}+b_1 \frac{s^2}{2}\,+ \right. \nonumber\\
   &+& \left. b_2 \left( \ln(s+1)-s+\frac{s^2}{2} \right)+\frac{s^3}{3}\right]
 \label{eq:masa}
\end{eqnarray}

\begin{equation}
\rho_{s} =\rho_{s}^{\rmn{NFW}}-\left[\frac{1}{3}+\frac{1}{2}b_{1}+
\left(\ln2-\frac{1}{2}\right)b_{2} \right] c^3 g(c) \bar{\rho}
\label{eq:rhos}
\end{equation}

Applying the condition (\ref{eq:mv1}) to the expression (\ref{eq:masa}) leads to Eq.~(\ref{eq:rhos}), which quantifies the influence of external parameters $b_1$ and $b_2$ on the value of $\rho_s$. In the case of $b_1=b_2=0$, which is equivalent to $f(s)=0$, we recover the modified NFW characteristic density $\rho_s^{\rmn{mod}}$.

An example of this new fitting formula for Milky Way size haloes is shown in Figure~1. Here we compare different approximations to fit the numerical density profile. In order to reduce the range of variation along the $y$-axis, we plot the function $ \left(\rho/\bar{\rho}\right) \cdot (r/R_{\rmn{vir}})^{2} $ instead of just $ \rho/\bar{\rho} $. The relative errors for each one of them are also displayed. The numerical profile given by cosmological simulations (dotted line) is compared to the NFW model (Eq.~\ref{eq:nfw}), the modified NFW model (NFW plus the mean matter density of the Universe, Eq.~\ref{eq:nfwmod}) and our model (Eq.~\ref{eq:ourmodel}). The fit in the inner part is quite similar for all the models as expected. There is an obvious discrepancy beyond the virial radius between the NFW fit, the modified NFW approach, and the data from numerical simulations. Our model helps to alleviate this disagreement, providing a reasonable fit going from the inner regions to well beyond the virial radius up to $10R_{\rmn{vir}}$.

\begin{table*}
\caption {Results of the fitting of our approximation to average halo density profiles in different mass bins. The first column sets the name of the bin, the second shows the number of distinct haloes in each mass range, the third corresponds to the mass range in logarithmic scale, the fourth is the mean virial mass in this range, the fifth the concentration in the NFW model, the sixth, seventh and eighth show the best-fitting values of the parameters of our approximation. The ninth and tenth columns indicate the position of the local extrema in the numerical density profiles.}
\begin{tabular}{cccccccccc}
\hline
      Name &     Number & Mass range &     $<M_{\rmn{vir}}>$ &      $c_{\rmn{NFW}}$ &          $c$ &         $b_1$ &         $b_2$ &       $s_{\rmn{max}}$ &       $s_{\rmn{min}}$ \\

           &   of haloes & $\log_{10}$($M_{\rmn{vir}} [h^{-1}M_{\odot}]$) &     $[h^{-1}M_{\odot}]$ &            &            &            &            &            &            \\
\hline
         H1 &       3512 & [ 11.5 , 12.0 ) & $5.35\times10^{11}$ &      12.99 &      12.43 &     -19.00 &      24.17 & 0.080 & 2.623 \\

         H2 &       4591 & [ 12.0 , 12.5 ) & $1.64\times10^{12}$ &      10.84 &      10.15 &     -23.43 &      30.01 & 0.097 & 2.661 \\

         H3 &       1719 & [ 12.5 , 13.0 ) & $5.08\times10^{12}$ &       9.40 &       8.73 &     -27.83 &      34.57 & 0.112 & 2.750 \\

         H4 &       4894 & [ 13.0 , 13.5 ) & $1.58\times10^{13}$ &       7.99 &       7.19 &     -33.87 &      41.61 & 0.134 & 2.797 \\

         H5 &       1509 & [ 13.5 , 14.0 ) & $4.82\times10^{13}$ &       7.52 &       6.82 &     -37.11 &      45.18 & 0.141 & 2.820 \\

         H6 &        384 & [ 14.0 , 14.5 ) & $1.48\times10^{14}$ &       6.75 &       6.12 &     -44.20 &      52.95 & 0.155 & 2.872 \\

         H7 &         68 & [ 14.5 , 15.0 ) & $4.47\times10^{14}$ &       5.76 &       5.26 &     -48.94 &      58.04 & 0.177 & 2.954 \\
\hline
\end{tabular}
\end{table*}

The description of the DM distribution from small to large distances from the centre of a halo presents an added relevance as it is related to the halo-dark matter correlation function $\xi_{\rmn  {Halo-DM}}= \frac{\left\langle\rho(r)\right\rangle -\bar{\rho}}{\bar{\rho}} $ (e.g. \citealt{Ha08}). Therefore, our approximation for the density profiles in the outer regions of DM haloes is also showing information about this correlation function. In Figure~2 we display $\xi_{\rmn  {Halo-DM}}$ for different bins in the mass range covered in this paper (see Section~3). In this plot we can clearly distinguish between the one-halo term, which corresponds to the halo density profile, and the two-halo term, which corresponds to the distribution of matter in its outskirts. Our approximation provides a reasonable description in both regions.

\begin{figure}
\includegraphics[width=0.5\textwidth]{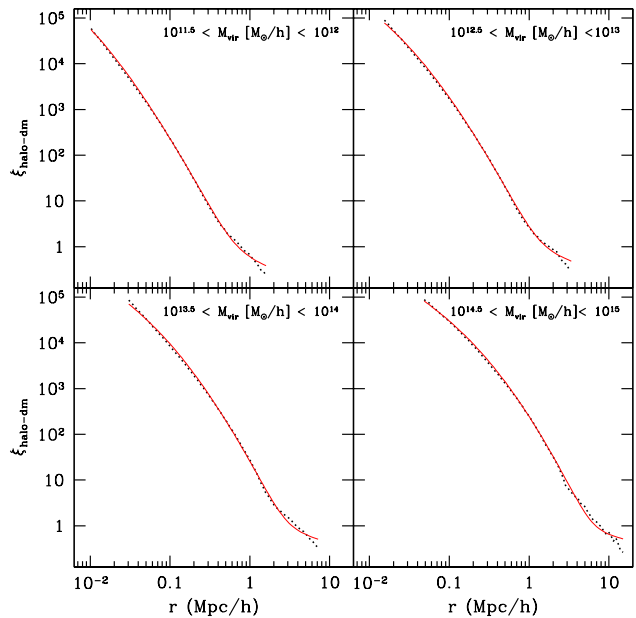}
\caption{Halo-dark matter correlation function $\xi_{\rmn{Halo-DM}}$ for average haloes in four different mass bins. The solid line represents our approximation whereas the dotted curves are the results from numerical simulations.}
\end{figure}

\section{Fit to simulated density profiles}
\label{sec:ajustes}
   To carry out our analysis we have used several high-resolution cosmological simulations. The initial conditions of these simulations have been set up from the power spectrum corresponding to the $\Lambda$CDM model with the following values of the cosmological parameters: $\Omega _{m} $= 0.3, $\Omega _{\Lambda} $= 0.7, $h$ = 0.7 and $\sigma _{8} $= 0.9. We used three different boxes whose size is 80, 120 and 250 $h^{-1}$Mpc. The mass resolution in these boxes is $3.14 \times 10^8$, $1.06 \times 10^9$ and $9.59 \times 10^9 h^{-1} M_{\odot} $ respectively, and the space resolution is 1.2, 1.8 and 7.6 $h^{-1}$kpc, respectively. The final output of the simulation is obtained by tracking the evolution of $512^3$ particles in each box to $z = 0$, using the $N$-body code ART \citep{Kr97}. The collapsed DM haloes are detected using Bound Density Maxima halofinder (BDM).

   We use mean halo density profiles averaged over seven different virial mass bins ranging from $5.35\times10^{11} h^{-1}M_{\odot} $ to $4.47\times10^{14} h^{-1} M_{\odot} $. There is a total of 17,220 haloes in all simulation boxes but we constrain this sample with the following selection criterion in order to select distinct haloes: the centre of all selected haloes must be further than the virial radius of every halo with higher mass. This criterion reduces the number of haloes to 16,679 (see \citealt{Cu07} for details on this halo sample).

   Figure~3 shows the best fit provided by our approximation to some average density profiles from simulations. In this plot we can see that there are deviations of the order of 15--20\% at the position $s_{\rmn{max}}$ of the local maximum of $\rho(r)\cdot r^2$. This deviation is inherited from the NFW model, where the fit to numerical density profiles usually show a noticeable disagreement at $r=r_s$. However, the transition from the internal to external regions at 2--3$R_{\rmn{vir}}$ of the halo displays a similar difference due to the fact that our simple model is not able to reproduce the steepest region in the density profile. In Table~1 we give the results of the best fit of our model to the simulated average halo density profile for the seven different mass bins.

\begin{figure*}
  \includegraphics[width=1.0\textwidth]{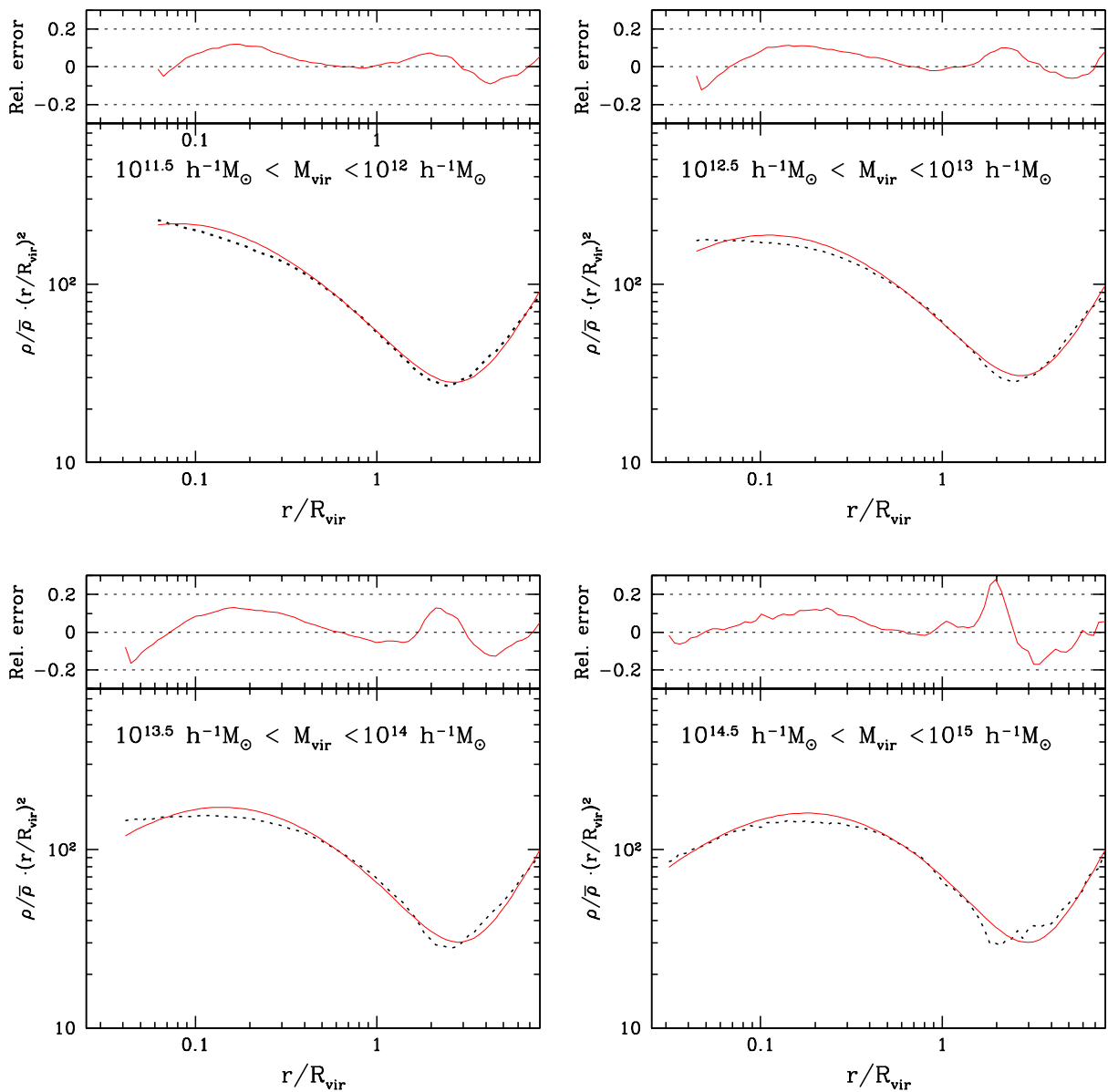}\\
  \caption{Average density profiles in four different halo mass ranges. The solid line represents the best-fit of our approximation to the numerical density profile (dotted line). The deviation around $r=r_s$ is inherited to the NFW profile, which is about the same amount of the deviation at the transition to the outer regions of the halo at 2--3$R_{\rmn{vir}}$.}
\end{figure*}

\begin{figure}
  \includegraphics[width=0.5\textwidth]{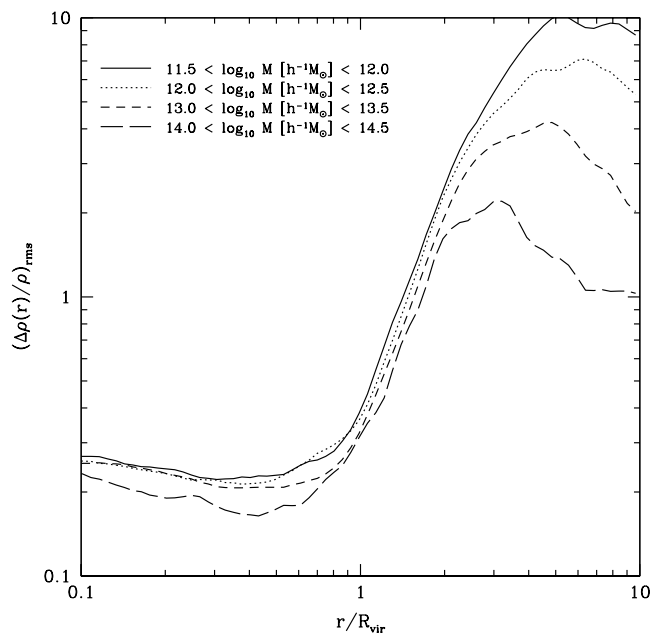}\\
  \caption{Halo-to-halo variation of DM density profiles in four different mass bins. Whereas the fluctuations compared to the magnitude of the density profile are small below $1R_{\rmn{vir}}$, they become very large around $r\gtrsim2R_{\rmn{vir}}$. At large distances, this scatter is higher for decreasing halo mass. This effect is not an artifact of the different behaviour of mean density profile at large distances, but the intrinsic scatter due to halo environment.}
\end{figure}

\begin{figure}
  \includegraphics[width=0.5\textwidth]{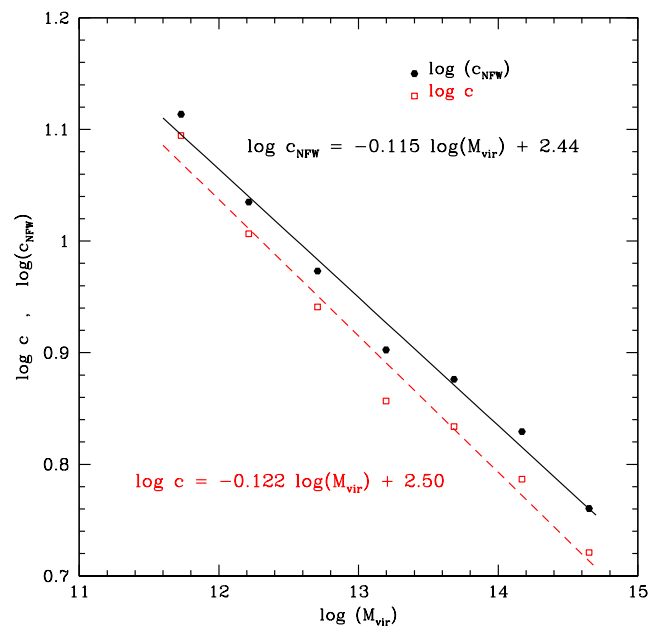}\\
  \caption{The correlation between the average virial mass in the mass bins under study and the concentration parameters, i.e. both the usual NFW concentration (solid circles) and the $c$ in our approximation (open circles). Solid and dashed lines represent the best fit to a power-law for $c_{\rmn{NFW}}(M_{\rmn{vir}}$ and $c(M_{\rmn{vir}}$, respectively.}
\end{figure}
\begin{figure}
  \includegraphics[width=0.5\textwidth]{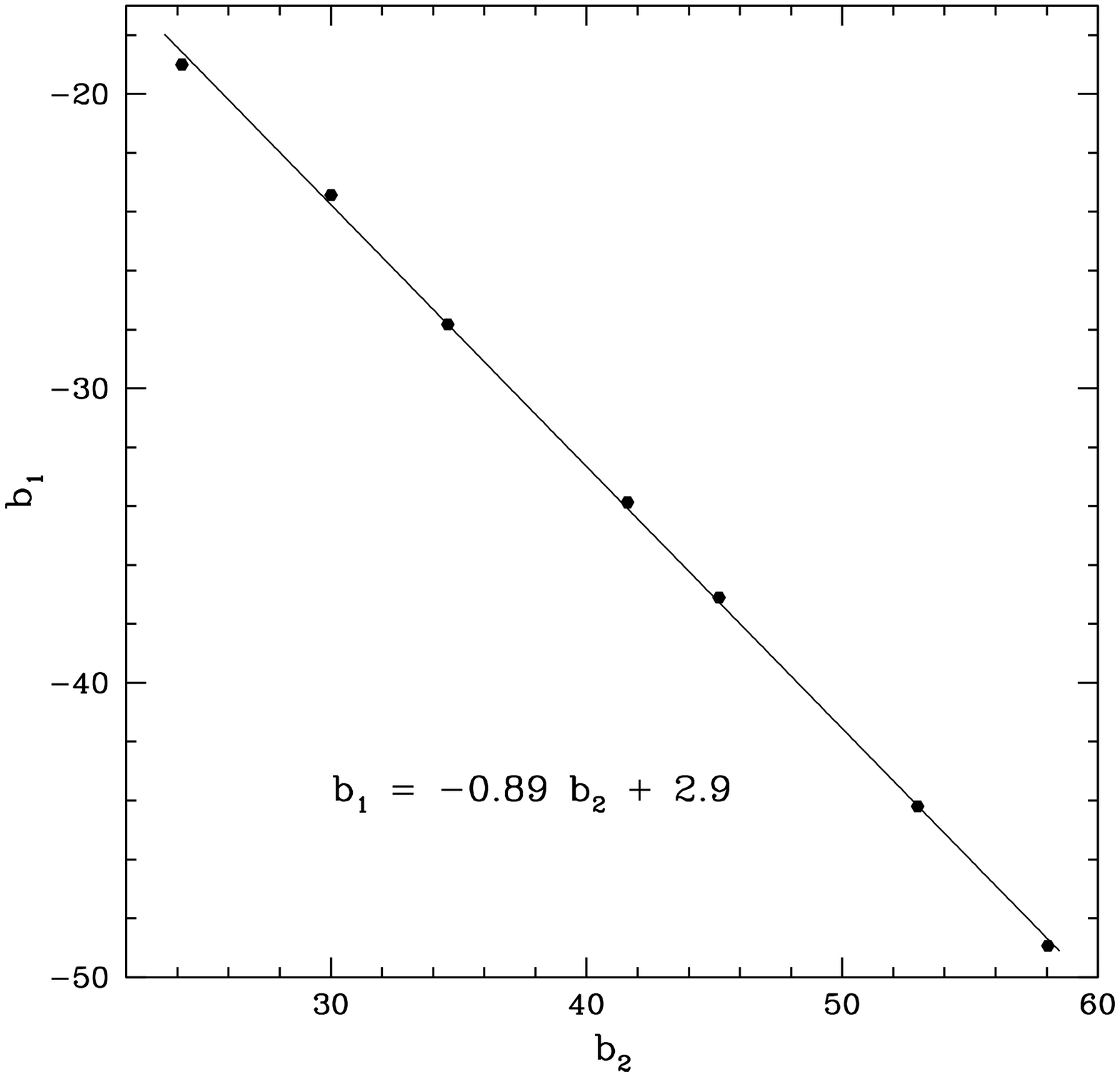}\\
  \includegraphics[width=0.5\textwidth]{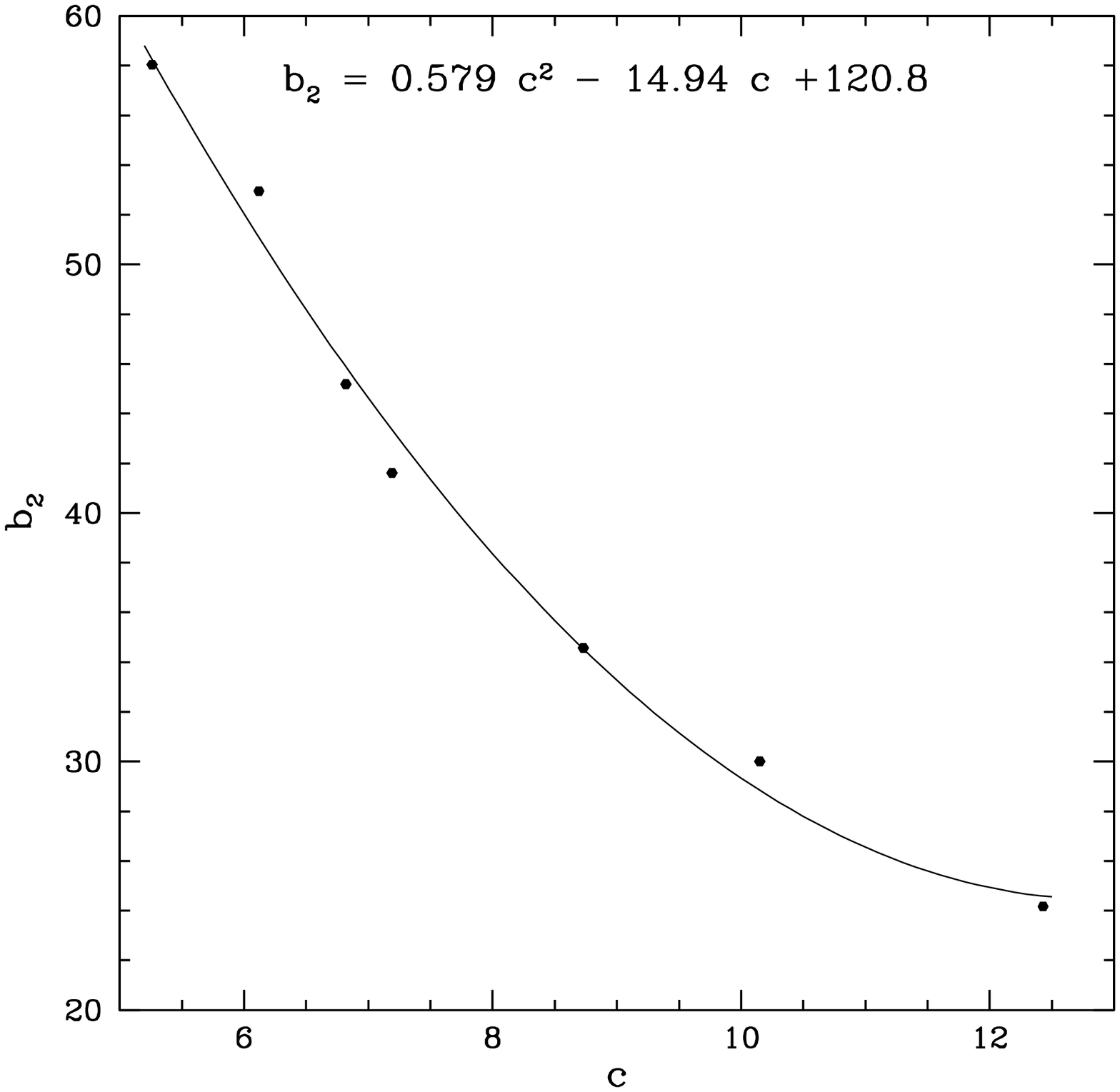}\\
  \caption{\textit{Top panel}: correlation between the parameters $b_1$ and $b_2$ in our approximation. There is a tight linear correlation between both parameters which describe the outskirts of the density profile, so our model gains simplicity. \textit{Bottom panel}: correlation between the parameters $b_2$ and $c$ in our approximation. Although a linear regression fails in order to find a correlation of this two parameters, the quadratic regression works reasonably well.}
\end{figure}

It is interesting to keep in mind the halo-to-halo variation of DM density profiles. This has been shown for example in \citet{Pr06}. In Figure~4 we show the root mean square in $\rho(r)$ normalized to the density profile itself. Although only small variations (at the level of 20--30\%) are present inside $R_{\rmn {vir}}$ among different haloes, the density profiles show fluctuations larger than twice the density profile itself beyond $2R_{\rmn {vir}}$. Haloes corresponding to small masses present a larger scatter.

\begin{table*}
\caption{Results from the regression analysis. The first two columns display the name of the correlated variables (dependent and independent variable respectively). In the remaining columns,  the value of Pearson's correlation coefficient, the coefficients of the regression equations and their standard deviations are shown. In the first four rows a linear regression is performed between both variables $(y=mx+n)$, whereas in the last two rows a quadratic relationship is shown $(y=px^2+mx+n)$.}
\begin{tabular}{ccccccccc}
\hline
         y &          x &         $r^2$ &        $n$ &       $\sigma_n$ &          $m$ &         $\sigma_m$ &         $p$ &        $\sigma_p$ \\
\hline
      $c_{\rmn{NFW}}$ &         $c$ & 0.9985 & 0.66 & 0.15 & 1.00 & 0.02 &      -      &      -      \\

     $\textrm{log}(c_{\rmn{NFW}})$ & $\textrm{log}(M_{\rmn{vir}})$ &   0.9857 & 2.44 &  0.08 & -0.115 & 0.006 &     -       &     -       \\

     log c & $\textrm{log}(M_{\rmn{vir}})$ &      0.9788 & 2.50 & 0.11 & -0.122 & 0.008 &     -       &      -      \\

        $b_1$ &         $b_2$ & 0.9994 &  2.9 & 0.4 & -0.890 & 0.0010 &     -       &     -       \\

        $b_2$ &          $c$ & 0.9904 & 121 & 8 & -15 & 2 & 0.58 &  0.11 \\

        $b_1$ &          $c$ & 0.9905 & -107 & 8 & 14 & 2 & -0.56 & 0.10 \\
\hline
\end{tabular}
\end{table*}

A regression analysis between the parameters in our approximation allows us to find that there are strong correlations among them. These results are shown in Table~2. An interesting correlation occurs between the NFW concentration (hereafter $c_{\rmn{NFW}}$) and the parameter $c$ in our model. This tight relation motivates the search for an interpretation of this parameter. We recall that in the NFW model the concentration is related to the radius in which the function $\rho(r)\cdot r^{2} $ presents its maximum, i.e. $s_{\rmn{max}}=1/c_{\rmn{NFW}} $. It is tempting to explore whether the parameter $c$ in our model admits a similar interpretation. The condition $d/ds \left (s^{2} \cdot  \rho/\bar{\rho} \right)=0 $ is a sixth degree polynomial equation with two real and positive solutions, $s_{\rmn{max}}$ and $s_{\rmn{min}}$, corresponding to the local maximum and minimum of this function respectively. The numerical solutions for $s_{\rmn{max}}$ and $s_{\rmn{min}}$ are shown in the last two columns of Table~1. The correlation between $c_{\rmn{NFW}}= 1/s_{\rmn{max}}$ and $c$ turns out to be:

\begin{equation}
\frac {1}{s_{\rmn{max}}}=0.967(\pm 0.004)\cdot c+0.53(\pm 0.03)
\label{eq:csmax}
\end{equation}
with a Pearson correlation coefficient of $1-r^2=7.3 \times 10^{-5} $ showing this evident connection. This means that our parameter $c$, which is no longer related to the position of the local maximum of $\rho(r) r^{2}$ due to the addition of $f(s)$, still retains its original interpretation, with only a small variation in its value. The function $f(s)$ has therefore very little influence in the inner regions, i.e. in the NFW term of our approximation.

The correlation between the NFW concentration and the virial mass is already well-known (e.g. \citealt{Ma07}). Therefore, the relationship between our parameter $c$ and $M_{\rmn{vir}}$ is straightforward as shown in Figure~5 for our sample average halo profiles. This correlation is extremely useful as it will allow us to derive simple mathematical expressions for the outer density profile in terms of the parameter $c$, which would be more complicated if they are written in terms of the virial mass. However, it is important to remark that this choice is just for our convenience, as we know that the parameter $c$ has no effect in the outer halo regions.

Something more surprising arises when we examine the relation among the fitted parameters $b_1$, $b_2$ and $c$ (as it can be readily seen from the last two rows in Table 2). This allows us to remove two parameters from our density model (Eq.~12) as $b_1$ and $b_2$ can be expressed in terms of $c$ with little scatter.

Thus, taking into account that the parameters $\rho_s$, $b_1$ and $b_2$ are a function of $c$, expression (\ref{eq:ourmodel}) turns into:
\begin{eqnarray}
\frac{\rho(s)} {\bar{\rho}} & = & \frac{\rho_s(c)/\bar{\rho}} {cs \ (cs+1)^{2}} +\frac{b_1(c)}{s}+\frac{b_2(c)}{s+1} +1 =\nonumber\\
& = & \underbrace{\frac{\rho_s/\bar{\rho}} {cs \ (cs+1)^{2}}}_{\rmn{internal}}+\underbrace{\frac{(b_1+b_2)s+b_1}{s(s+1)}}_{\rmn{middle}}+\underbrace{1}_{\rmn{external}}
\label{eq:threeparts}
\end{eqnarray}
where $\rho_s(c)/\bar{\rho}$, $b_1(c)$ and $b_2(c)$ are no longer free parameters but fixed in terms of $c$. This dependence is determined by the expression (\ref{eq:rhos}) and the values shown in the last two rows in Table~2, i.e.,
\begin{equation}
\begin{array}{ccrrr}
 \rho_s / \bar{\rho} (c) & = & \,0.167\,c^2 & -4.12\,c & +143.4 \nonumber\\
 b_2(c) & = & \, 0.579\,c^2 & -14.94\,c & +120.8 \nonumber\\
 b_1(c) & = & -0.558\,c^2 & +14.02\,c & -107.4 \nonumber
 \label{eq:parc}
\end{array}
\end{equation}
It is important to realize that Eq. (\ref{eq:threeparts}) has been grouped in such a way that each term is related to a region of the halo density profile (internal, middle and external, respectively).

After the elimination of $b_1$ and $b_2$, we obtain again the best-fit for the only free parameter $c$, taking into account that now the values of $b_1$ and $b_2$ are fixed and can be obtained using Eq.~(\ref{eq:parc}). We note that Eq.~(\ref{eq:parc}) implicitly includes the correlation between $c$ and $M_{\rmn{vir}}$, as the use of the parameter $c$ instead of $M_{\rmn{vir}}$ makes this expression much simpler. We have chosen this form for our mere convenience although $c$ has nothing to do with the outer halo regions. The results are given in Table~3, where the fourth and fifth column allows a comparison with Table~1. Only small differences about few per cent are present between the best-fit values for $b_1$ and $b_2$ in Table~1 and the new results for the best-fit value of $c$ in Table~3.
\\
\begin{table}
\begin{center}
\caption {Results of the constrained fitting of our approximation to halo density profiles. As opposed to Table~1, here $c$ is the only free parameter, whereas the values of $b_1(c)$ and $b_2(c)$ are obtained using Eq.~(\ref{eq:parc}) for the best-fit value of $c$. We include the fourth and fifth column to allow direct comparison with Table~1.}
\begin{tabular}{cccccc}
\hline
      Name &   $M_{\rmn{vir}}$   &     c &        $b_1(c)$ &        $b_2(c)$ \\

           &  ($M_{\odot}$/h) &  &       &            &            \\
\hline
         H1 &  $5.35 \times 10^{11}$  &  12.42 &   -19.32 &    24.60 \\

         H2 &  $1.64 \times 10^{12}$  &  10.12 &   -22.66 &    28.95 \\

         H3 &  $5.08 \times 10^{12}$  &   8.77 &   -27.37 &    34.35 \\

         H4 &  $1.58 \times 10^{13}$  &   7.12 &   -35.89 &    43.82 \\

         H5 &  $4.82 \times 10^{13}$  &   6.80 &   -37.89 &    46.02 \\

         H6 &  $1.48 \times 10^{14}$  &   6.17 &   -42.17 &    50.70 \\

         H7 &  $4.47 \times 10^{14}$  &   5.26 &   -49.12 &    58.27 \\
\hline
\end{tabular}
\end{center}
\end{table}

Apart from the discrepancies already shown in Figure~3, we also note that the slope in the density profiles from the simulations tends to be shallower as compared to our approximation starting around 8 virial radii. We therefore address the issue of the range of validity of our approximation. In order to study this, we will use a density profile which extends up to nearly 30 virial radii (see Fig.~7).

\begin{figure}
\includegraphics[width=0.5\textwidth]{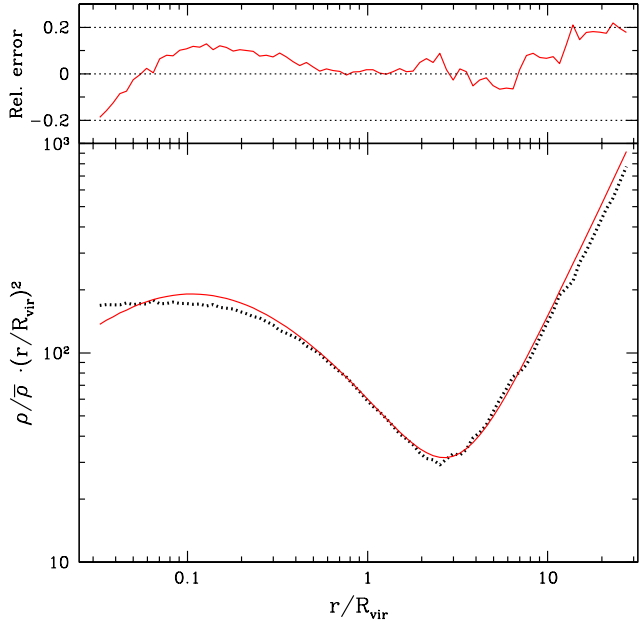}
\caption{The density profile of an average halo with $M_{\rmn{vir}}\simeq 5 \times 10^{12} \ h^{-1}M_{\odot}$ up to nearly $30R_{\rmn{vir}}$. Although our approximation implements the right trend at large distances from the halo centre, residuals of the order of $1/r$ are amplified since the density profile is multiplied here by $r^2$.}
\end{figure}

At these very large distances we are able to analyse the asymptotic behaviour of both numerical and approximated density profiles. To this aim, it is more useful to use the function $y=(\rho/\bar{\rho}) \ s$  since we can study its oblique asymptote. If we multiply Eq.~(\ref{eq:ourmodel}) by $s$ we obtain that the asymptotic behaviour is:
\begin{equation}
y=s+b_{1}+b_{2}
\label{eq:asymt}
\end{equation}

The leading term at large distances corresponds to the trend of the density profile to have the same value as the mean matter density in the Universe, but there is also a dependence on the coefficients $b_{1}$ and $b_{2}$. In Figure~8 we can also see that this dependence makes our approximation slightly shifted by a constant value $b_1+b_2$ with respect to the data from cosmological simulations. This means that the $\mathcal{O}(1/r)$ terms in our approximation are still important in the range from 10 to $\sim 30R_{\rmn{vir}}$ just before entering the asymptotic regime where the density profile tends to the constant mean matter density $\bar{\rho}$.

\begin{figure}
\includegraphics[width=0.5\textwidth]{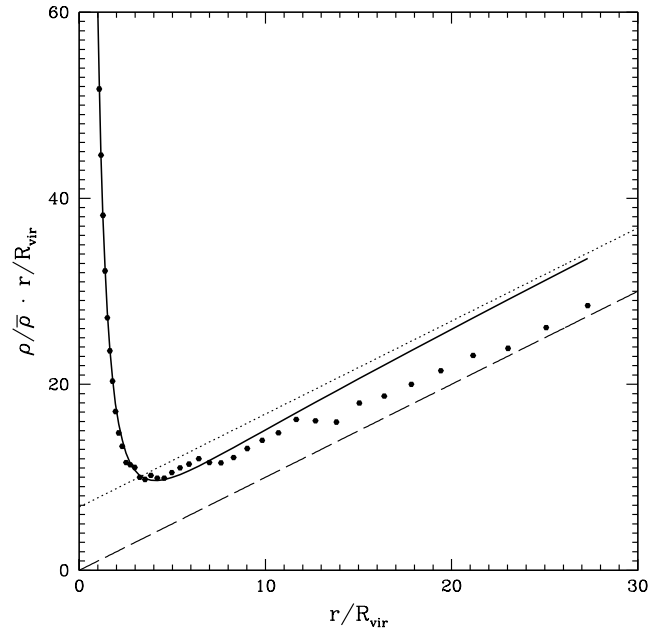}
\caption{The density profile $\rho/\bar{\rho}$ times $r/R_{\rmn{vir}}$ versus $r/R_{\rmn{vir}}$ for our approximation (solid line) and the data from numerical simulations (solid circles). The dashed and the dotted lines represent the oblique asymptotes which correspond to our fit and the simulation data, respectively. The vertical shift between both asymptotes shows the presence of $1/r$ errors in our approximation.}
\label{fig:14}
\end{figure}

This shows the limitations of our approximation: while it provides a reasonable fit up to 10 virial radii, the goodness of the fit is not so good at larger distances. This is mainly due to the simplicity of the approximation, which is one of the premises in our model. Of course, we can always add more rational terms to our function $f$ but this would complicate the model and the interpretation of the different terms. However, the density at distances larger than $10R_{\rmn{vir}}$ is similar to the mean matter density, e.g. $\rho(s>8)<1.5 \bar{\rho}$. With this in mind, we can calculate the enclosed mass in a sphere of a given radius, and compare with our approximation (see Figure~9). We can see that even comparing with the modified NFW density profile (i.e. NFW plus the mean matter density in the Universe), our approximation is a significant improvement regarding the enclosed mass. While the modified NFW profile can overestimate the enclosed mass by 5\% at $3R_{\rmn{vir}}$, and underestimate it by 10\% at $8R_{\rmn{vir}}$, our approximation keeps this uncertainty below 2\% even at $10R_{\rmn{vir}}$, which makes our simple model more than enough for this application. On the contrary, the plain NFW profile is not suitable beyond $4R_{\rmn{vir}}$ in order to estimate the enclosed mass, as the estimated mass falls short by more than a factor of two with respect to the data from the simulations at $10R_{\rmn{vir}}$ (see Figure~9).

\begin{figure}
\includegraphics[width=0.5\textwidth]{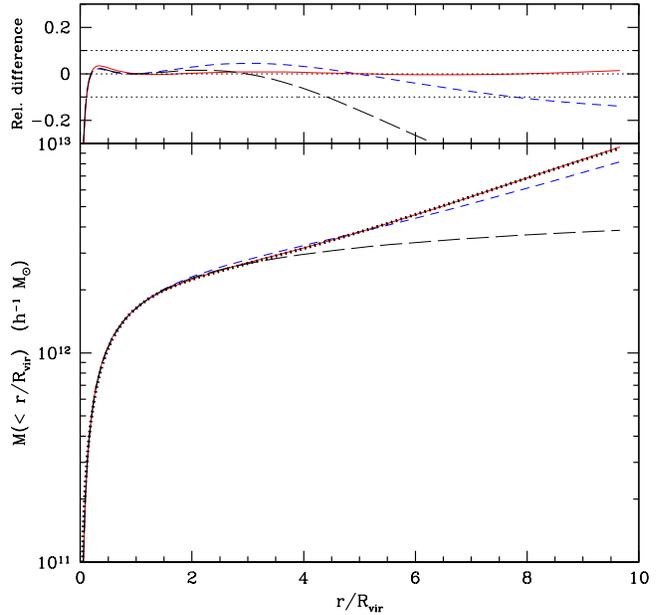}\\
\caption{Enclosed mass in a sphere of radius $r$ versus $r/R_{\rmn{vir}}$ for the average halo with $M_{\rmn{vir}}\simeq 10^{12}h^{-1}M_{\odot}$. The dotted curve represents the simulation data, long dashed and short dashed lines show the plain NFW and the modified NFW profile (i.e. NFW plus the mean matter density $\bar{\rho}$) respectively. The solid line is our approximation. Top panel represents the relative errors with respect to the simulation data.}
\end{figure}

\section{Application to gravitational lensing}

One of the most relevant experimental techniques for determining the distribution of dark matter is based on the gravitational lensing effect (for some specific applications, see e.g. \citealt{Ga03}, \citealt{Br05}, \citealt{Jo07}, \citealt{Ma06}, \citealt{Ma08}). In these observations the quantity in which we are interested is the tangential shear $\gamma_t$, which describes the image elongation perpendicular to the line connecting the image from the source and the distorting lens. This quantity is related with the surface mass density $\Sigma$ (projected on the lens plane, which is perpendicular to the trajectory of the incoming ray), through the following formula:

\begin{equation}
\gamma_t(R)=\frac{\Delta \Sigma} {\Sigma_{crit}}=\frac{\bar{\Sigma}(<R)-\Sigma (R)}{\Sigma_{crit}}
\end{equation}
where $R$ is the radial coordinate projected on the plane of the lens (impact parameter). $\bar{\Sigma}(<R)$ is the mean surface density enclosed in a radius $R$, and $\Sigma_{crit}$ is the critical surface density, which depends from the distance between the lens and the source, i.e.,

\begin{equation}
\Sigma_{crit}=\frac{c^2}{4 \pi G} \, \frac{D_S}{D_L D_{LS}}
\end{equation}
where $D_S$, $D_L$ and $D_{LS}$ are the distance to the source, the distance to the lens, and the distance between the source and the lens, respectively.

In fact, the quantity which can be observed is the shear in units of critical density $\Delta \Sigma (R)$. We can derive its expression using our approximation for the dark matter halo density profile.

The surface mass density is given by the following formula, i.e. the projection of the volume density $\rho(r)$ in the line of sight:
\begin{equation}
\Sigma (R)=2 \int_{R}^{+\infty} \rho (r) \frac {r}{\sqrt{r^2-R^2}} \ dr .
\end{equation}
instead, the mean surface density is defined as follows:
\begin{equation}
\bar{\Sigma}(<R)=\frac{1}{R}\int_{0}^{R} \Sigma (R') \ dR' .
\end{equation}
For simplicity in our calculations, we use distances which are scaled to the virial radius ($s\equiv\frac{r}{R_{\rmn{vir}}}$; $S\equiv\frac{R}{R_{\rmn{vir}}}$). Therefore, the last two equations transform into:
\begin{equation}
\Sigma (S)=2R_{\rmn{vir}} \int_{S}^{+\infty}  \rho (s) \frac {s}{\sqrt{s^2-S^2}} \ ds
\end{equation}
\begin{equation}
\bar{\Sigma}(<S)=\frac{1}{S}\int_{0}^{S} \Sigma (S') \ dS'
\end{equation}

where we choose for $\rho(s)$ our approximation for the density profile given by Eq.~(\ref{eq:threeparts}), in which $b_1$ and $b_2$ are functions which depend on our parameter $c$ as explained in the previous section, i.e.,
\begin{equation}
\rho(s)=\underbrace{\frac{\rho_{s}}{cs\left(1+cs\right)^2}}_{\rho_{int}(s)}+ \underbrace{\frac{b_{1}(c)}{s}\bar{\rho}+\frac{b_{2}(c)}{s+1}\bar{\rho}}_{\rho_{mid}(s)}+ \bar{\rho}
\label{eq:3parts}
\end{equation}

Now we compute the tangential shear for this density profile. As we can separate the contributions from different terms in Eq.~(\ref{eq:3parts}) due to linearity of the integrals in the definition of the shear, we will calculate each term (i.e. the internal, the middle region and the contribution from background density) separately. The last term has associated a trivial contribution, because it is a known fact that for any homogeneous density field the tangential shear is null. Therefore, in the case of the background density we obtain $\Delta \Sigma_b(S)=0$.

\subsection{Internal tangential shear}

The first term in this equation is formally identical to the NFW model and the surface mass density $\Sigma$ for this model has already been determined by \citet{Ba96}:

\begin{equation}
\Sigma_{int} (S)=\frac {2\rho_s R_{\rmn{vir}}}{c} \int_{S}^{+\infty} \frac {ds}{(cs+1)^2\sqrt{s^2-S^2}}
\end{equation}

\begin{equation}
\Sigma_{int}(S)=\left \lbrace \begin{array} {ccc}\frac{2 \rho_s \ R_{\rmn{vir}}}{c} \ \frac{1-\phi(cS)}{c^2S^2-1} & \textrm{if} & S \ne \frac{1}{c} \\ \\ \frac{2 \rho_s \ R_{\rmn{vir}}}{3c} & \textrm{if} & S = \frac{1}{c} \end{array} \right.
\end{equation}\nonumber
where $\phi$ is given by:

\begin{equation}
\phi(x)=\left \lbrace \begin{array}{ccc} \frac{\textrm{atanh}\sqrt{1-x^2}}{\sqrt{1-x^2}} & 0<x<1 \\ \\ 1 & x=1 \\ \\ \frac{\textrm{arctan}\sqrt{x^2-1}}{\sqrt{x^2-1}} & x>1 \end{array} \right.
\end{equation}
On the other hand, we get from the calculation of the mean surface density:
\begin{equation}
\bar{\Sigma}_{int}(<S)=\frac {2 \rho_s R_{\rmn{vir}}}{c} \left \lbrace \begin{array} {cc} \int_{0}^{S} \frac{1-\phi(cS')}{c^2S'^2-1}dS' &  0<S<1/c \\ \\ 1/c & S=1/c \\ \\ \frac {1}{c}+\int_{1/c}^{S} \frac{1-\phi(cS')}{c^2S'^2-1}dS' &  S>1/c \end{array} \right.
\end{equation}
and hence we can write this in terms of $\phi$, i.e.:
\begin{equation}
\bar{\Sigma}_{int}(<S)=\frac{2 \rho_s \ R_{\rmn{vir}}}{c} \ \phi(cS) .
\end{equation}
Therefore the tangential shear $\Delta \Sigma_{int} (S)=\bar{\Sigma}_{int}(<S) - \Sigma_{int}(S)$ is:
\begin{equation}
\Delta \Sigma_{int} (S)=\frac{2 \rho_s R_{\rmn{vir}}}{c} \left \lbrace \begin{array} {cc} \frac{c^2 S^2 \phi(cS)-1}{c^2S^2-1} & S \ne 1/c \\ \\ 2/3 & S=1/c \end{array} \right.
\end{equation}

So, if we introduce for our convenience the function $\psi$ defined as follows:

\begin{equation}
\psi(x)=\left \lbrace \begin{array} {cc} \frac{x^2 \phi(x)-1}{x^2-1} & x \ne 1 \\ \\ 2/3 & x=1 \end{array} \right.
\end{equation}
We can simply write the tangential shear for the internal region in this form:
\begin{equation}
\Delta \Sigma_{int}(S)=\frac{2 \rho_s \ R_{\rmn{vir}}}{c} \ \psi(cS)
\end{equation}

\subsection{Middle tangential shear}

The second and third terms in the equation (\ref{eq:3parts}) represent the contribution from the intermediate region of the density profile to the tangential shear. For $\rho_{mid}(s)=\left( \frac{b_1(c)}{s}+\frac{b_2(c)}{s+1} \right) \bar{\rho}$ the surface density is:

{\setlength\arraycolsep{2pt}
\begin{eqnarray}
\Sigma_{mid}(S) & = & 2 R_{\rmn{vir}} \bar{\rho} \int_{S}^{\infty}\frac{(b_1(c)+b_2(c)) \ s \ ds}{(s+1)\sqrt{s^2-S^2}} \nonumber\\
& = & 2 R_{\rmn{vir}} \bar{\rho} \bigg \lbrack (b_1+b_2)\lim_{\xi \to +\infty} \ln \frac{\sqrt{\xi^2-S^2}+\xi}{S}+\phi(S) \bigg \rbrack \nonumber
\end{eqnarray}}

where $\phi(s)$ is defined in the same way as above, and for the mean surface density we get:

{\setlength\arraycolsep{2pt}
\begin{eqnarray}
\bar{\Sigma}_{mid}(<S) & = & 2 R_{\rmn{vir}} \bar{\rho} \bigg \lbrack (b_1+b_2)+ \nonumber\\
& +& (b_1+b_2) \lim_{\xi \to \infty} \ln \frac{\sqrt{\xi^2-S^2}+\xi}{S}+ \frac{b_1}{S} \int_{0}^{S}\phi(S')\ dS' \bigg \rbrack \nonumber
\end{eqnarray}}
 The tangential shear from the intermediate region is therefore:
\begin{equation}
\Delta \Sigma_{mid}(S)=2 R_{\rmn{vir}} \bar{\rho} \bigg \lbrack b_1+b_2 \bigg ( 1+\phi(S)- \frac{1}{S}\int_{0}^{S} \phi(S')dS' \bigg ) \bigg \rbrack
\end{equation}
Unfortunately there is no primitive function for $\phi$ which could be written in terms of elementary functions. Nevertheless, it is possible to find a fitting function for it. We will fit $\frac{1}{S}\int_{0}^{S} \phi(S')dS'-\phi(S)$ over the interval $s\in [0,10]$ using the function $\frac{1}{1+ \left (x/a \right)}$. The result is:
\begin{equation}
\frac{1}{S} \int_{0}^{S} \phi(S')dS'-\phi(S) \approx \frac{1}{1+ \left ( \frac{S}{4.8} \right)} .
\end{equation}

We are now in a position to write down the complete expression of the tangential shear in units of the critical density for our model, i.e.,
\begin{equation}
\Delta\Sigma(S)=2 R_{\rmn{vir}}\bar{\rho} \bigg \lbrack \frac{\rho_s}{\bar{\rho}\ c} \psi (cS)+b_1+b_2 \left( \frac{S}{4.8+S} \right) \bigg \rbrack .
\end{equation}

We can now compare this tangential shear with that one given by the NFW model (which is formally identical to $ \Delta \Sigma_{int}(S)$, except that we have to replace $\rho_s$ and $c$ by $\rho_s^{\rmn{NFW}}$ and $c_{\rmn{NFW}}$ respectively):

\begin{equation}
\Delta \Sigma_{\rmn{NFW}}(S)=\frac{2 \rho^{\rmn{NFW}}_s \ R_{\rmn{vir}}}{c_{\rmn{NFW}}} \ \psi(c_{\rmn{NFW}}S)
\end{equation}

In Figure~10 we show the relative difference between $\Delta \Sigma_{\rmn{NFW}}(S)$  (which does not account for external region of the halo) and $\Delta\Sigma(S)$ as given by our model for the different halo mass bins. The differences are small and always below 4\%. They are not appreciable at the level of current experimental sensitivity. As expected, the contribution from the matter distribution in the outer regions of DM haloes is small, but for the first time it has been estimated to what extent it may have influence in the measurement of the tangential shear. Besides, we find an obvious trend with halo mass: the difference between both $\Delta \Sigma_{\rmn{NFW}}(S)$ and $\Delta\Sigma(S)$ is higher for most massive haloes.
\begin{figure}
\includegraphics[width=0.5\textwidth]{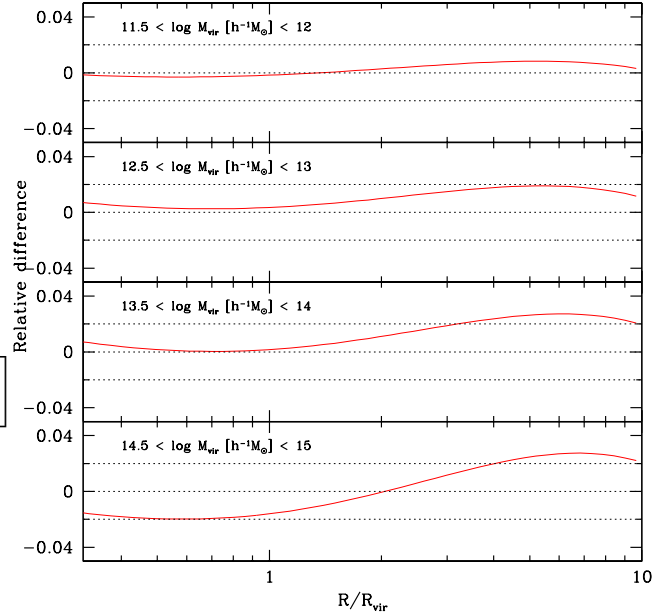}\\
\caption{The relative difference between the tangential shear (in units of critical surface density) for NFW approximation and our model, $(\Delta \Sigma_{\rmn{NFW}}(S) - \Delta\Sigma(S))/\Delta\Sigma(S)$, in four different mass ranges.}
\end{figure}

\section{Discussion and Conclusions}

In this paper we have presented a simple approximation for the DM density profiles of haloes with masses ranging from $10^{11.5}$ to $10^{15.0}h^{-1}M_{\odot}$, which is valid even beyond the virial radius up to $\sim 10R_{\rmn{vir}}$. The expression we are suggesting here is $\rho(s)=\frac{\rho_{s}}{cs\left(1+cs\right)^2}+\left(\frac{b_{1}}{s}+\frac{b_{2}}{s+1}+1\right)\bar{\rho}$, where $s=r/R_{\rmn{vir}}$. This approximation is an extension of the NFW formula but includes two additional parameters. We have shown that these parameters are very well correlated with the virial mass, so that the density profile is effectively just a function of $M_{\rmn{vir}}$. Other approximations found in the literature will prove to be useful to describe, with high accuracy, the halo density profile up to the virial radius. Yet, most of them fail when they are applied to fit the outer regions, where the mass predicted by extrapolation of these profiles is far below the actual mass in these regions.

The description of DM distribution far away from the halo centre is especially interesting. In particular, the halo--dark matter correlation function is related in a straightforward way to the \textit{average} density profile. Although the transition from the one-halo term to the two-halo term in this correlation function is present before $10R_{\rmn{vir}}$, our approximation has turned out to be a reasonable description of the DM distribution even at these distances. In order to build the numerical mean density profiles, we averaged over many hundreds of haloes from high-resolution cosmological simulations so that the profiles corresponding to most of our mass bins are entirely unaffected by statistics. This procedure for averaging density profiles is analogue to the stacking method, used in observational studies like the one by \citet{Ma08} to infer the density profile of a cluster of galaxies. This similarity is useful for the comparison of the results from cosmological simulations to the real data.

This parametrization for the average density profile is accurate to within 10--15\% in the range from $0.05R_{\rmn{vir}}$ to $10R_{\rmn{vir}}$. There are two main discrepancies from the numerical density profile which have a different origin: whereas the overestimation around $r=r_s$ is inherited from NFW profile, the overestimation just beyond the virial radius suggests that our model is not able to reproduce the steepest region. This steep region just outside $R_{\rmn{vir}}$ is more pronounced for most massive haloes, suggesting a depletion of the halo outskirts due to dark matter infall (\citealt{Pr06}, \citealt{Cu07}). The presence of our additional terms with respect to the NFW formula has only a very small influence on the inner regions of the density profile, so that our approximation can also be considered as an extension of the NFW profile. At larger distances our model shows deviations around 20\% in the range 10--30$R_{\rmn{vir}}$ just before entering the asymptotic regime. These deviations are caused by our additional $(r/R_{\rmn{vir}})^{-1}$ terms which improve the fit in the interesting region below $10R_{\rmn{vir}}$ where the density is much higher. In any case, we must remark that our approximation implements the correct asymptotic behaviour: the density profile tends to the asymptotic value of the mean matter density of the Universe $\bar{\rho}$.

The cumulative mass inside a sphere of a given radius is underestimated by more than 50\% at $10R_{\rmn{vir}}$ by the NFW formula. On the contrary, it is much better approximated (to within 12\%) when the NFW profile is modified by addition of the mean matter density, although with our model the difference with numerical density profiles is reduced even up to 1\% in the range 1--9$R_{\rmn{vir}}$. This is especially interesting for new measurements of the enclosed mass beyond virial radius in X-ray clusters \citep{Ge08}, where plain NFW is still used even at $r>1R_{\rmn{vir}}$. While current observations cannot distinguish between modified NFW and our approximation, in the near future they should be able to find the need for adding the mean matter density term to the density profile.

We have also presented an application for our approximation in the context of mass estimation using gravitational lensing effect. We derived expressions for tangential shear corresponding to different regions around the halo, which are in turn related to the different terms in our approximation. The contribution from the outer regions is small as compared to the contribution of the inner region, as expected. We calculated the difference between this tangential shear and the one derived from the NFW profile as a function of distance, showing that the inclusion of the outer regions produces a difference around 4\% . This small difference could provide an observational test for the validity of our approximation, which has been derived from the results of cosmological $N$--body simulations. Although present resolution of weak lensing experiments prevents us from drawing a robust conclusion, the stacking of different observations should prove that this approximation, which includes the contribution of external regions, is more realistic than most of the so far proposed density profiles, which do not account for them.\\

HT wants to thank the I.E.S. Jos\'e Cadalso for allowing him to combine this research with his teaching duties. FP, AJC and MASC thank the Spanish MEC under grant PNAYA 2005-07789 for their support. AK acknowledges support from NASA and NSF grants to NMSU. AJC acknowledges the financial support of the MEC through Spanish grant FPU AP2005-1826. MASC acknowledges the financial support of the CSIC through Spanish grant I3P.

\bibliography{mycites}

\bsp

\end{document}